**Accepted Version: Copy**

# Title: Li$_4$Ti$_5$O$_{12}$: A Visible-to-Infrared Broadband Electrochromic Material for Optical and Thermal Management

*Jyotirmoy Mandal, Sicen Du, Martin Dontigny, Karim Zaghib, Nanfang Yu\*, Yuan Yang\*.*

J. Mandal, S. Du, Prof. N. Yu\*, Prof. Y. Yang\*
Department of Applied Physics and Applied Mathematics, Columbia University,
Mudd 200, MC 4701, 500 W 120 Street, New York, NY 10027, U.S.A.
Email: yy2664@columbia.edu, ny2214@columbia.edu

M. Dontigny, Dr. K. Zaghib
IREQ−Institute Recherche d'Hydro-Québec
1800 Boulevard Lionel Boulet, Varennes, Québec J3X 1S1, Canada

***Abstract:*** *Broadband electrochromism from visible to infrared wavelengths is attractive for applications like smart windows, thermal-camouflage, and temperature control. In this work, the broadband electrochromic properties of Li$_4$Ti$_5$O$_{12}$ (LTO) and its suitability for infrared-camouflage and thermoregulation are investigated. Upon Li$^+$ intercalation, LTO changes from a wide band-gap semiconductor to a metal, causing LTO nanoparticles on metal to transition from a super-broadband optical reflector to a solar absorber and thermal emitter. Large tunabilities of 0.74, 0.68 and 0.30 are observed for the solar reflectance, mid-wave infrared (MWIR) emittance and long-wave infrared (LWIR) emittance respectively, with a tunability of 0.43 observed for a wavelength of 10 μm. The values exceed, or are comparable to notable performances in the literature. A promising cycling stability is also observed. MWIR and LWIR thermography reveal that the emittance of LTO-based electrodes can be electrochemically tuned to conceal them amidst* *their environment. Moreover, under different sky conditions, LTO shows promising solar heating and sub-ambient radiative cooling capabilities depending on the degree of lithiation and device design. The demonstrated capabilities of LTO make LTO-based electrochromic devices highly promising for infrared-camouflage applications in the defense sector, and for thermoregulation in space and terrestrial environments.*

**1. Introduction**

Electrochromism – a phenomenon in which certain materials reversibly switch colors with their redox state – is an increasingly familiar occurrence in today's world.[1] From use in smart windows and flexible displays, to demonstrated use in camouflaging,[2] electrochromic devices have found a wide range of applications. The term is commonly used to describe color changes in the visible spectrum. However, electrochromism occurs in other spectral ranges as well. Studies on visible electrochromism have been wide-ranging, with materials spanning from inorganic compounds (e.g. Prussian blue analogs and $WO_3$) to organic molecules and polymers (e.g. polyaniline) reported for their electrochromic properties.[3] In comparison, infrared electrochromism has been studied less extensively. However, electrochemical tunability of infrared emittance ($\epsilon$) from surfaces is highly attractive for optical and thermal applications. For example, tunable emittances in the mid-wave infrared (MWIR, wavelength ($\lambda$) ~ 3-5 μm) and long-wave infrared (LWIR, $\lambda$ ~ 8-13 μm) atmospheric transmission windows (ATWs) can be used for infrared-camouflage.[4] Another application is thermoregulation. Electrochromic devices with tunable emittances to control radiative cooling are already used in the space industry.[5] Likewise, visible-to-infrared electrochromic devices could also exhibit radiative heating or cooling on earth, and reduce electricity consumption and carbon footprint in residential and industrial settings. Materials exhibiting visible-to-infrared electrochromism are thus clearly attractive.

Research over recent decades has yielded several materials with visible-to-infrared electrochromic properties, such as poly(aniline) and $WO_3$.[6] Devices based on such materials typically show tunabilities of emittance ($\Delta\epsilon$) ≲ 0.3 in the MWIR and LWIR ATWs.[6a, 7] Two outliers are $H_xWO_3$-based devices (with $\Delta\epsilon_{MWIR}$ ~ 0.7 and $\Delta\epsilon_{LWIR}$ ~ 0.15-0.25) and poly(aniline)-based electrochromic skins (with super-broadband IR $\Delta\epsilon$ ~ 0.5).[5, 8] However, the use of aqueous $H^+$ and polymers could prevent their application at high temperatures. More importantly, the change in solar reflectance, which often determines radiative heat transfer under sunlight, is low for such designs. Therefore, electrochromic materials with high tunability in both visible and infrared spectral ranges remain sought-after.

Here, the authors report the electrochromic properties of $Li_4Ti_5O_{12}$ (LTO), a zero-strain material with highly reversible lithium ion ($Li^+$) intercalation, which exhibits tunable emissivity or reflectivity from visible to infrared wavelengths. LTO nanoparticles coated on aluminum (Al) are found to have an emittance that is electrochemically tunable by up to ~ 0.78 in the solar wavelengths (0.4-2.5 μm), and ~ 0.68 and ~ 0.30 in the MWIR and LWIR ATWs respectively, and show steady electrochromic behavior over a large number of cycles. The large tunability, along with its excellent cycle-life,[9] makes LTO attractive for infrared-camouflage and radiative thermoregulation.

## 2. Results and Discussion

### 2.1 The Electrochromic Properties of Nanostructured $Li_4Ti_5O_{12}$

LTO has been explored extensively as an anode material for Li-ion batteries due to its excellent cycle life, structural stability and fast charging/discharging rates as high as 100 C (corresponding to 36 s).[9-10] During lithiation, it transitions from a delithiated (DL) state ($Li_4Ti_5O_{12}$) to a lithiated (L) state ($Li_7Ti_5O_{12}$) (**Figure 1a**), which leads to a drastic change in electromagnetic properties. As shown in Figure 1b, $Li_4Ti_5O_{12}$ nanoparticles on Al have high reflectance ($R(\lambda)$) in the visible wavelengths, while $Li_7Ti_5O_{12}$ nanoparticles have high absorptance or emittance ($\epsilon(\lambda) = 1 - R(\lambda)$ for the opaque LTO on Al, according to Kirchhoff's rule, as detailed in Supporting Information, Section 1). Thermographs reveal that the contrasting optical properties of the two states extend well into the infrared (Figure 1b), and measurements of spectral reflectance $R(\lambda)$ confirm that an appreciable tunability > 0.4 stretches from $\lambda$ of 0.4 μm to 11 μm (Figure 1c), i.e., the electrochromic performance is super-broadband, and stretches across the solar, MWIR and LWIR wavelengths.

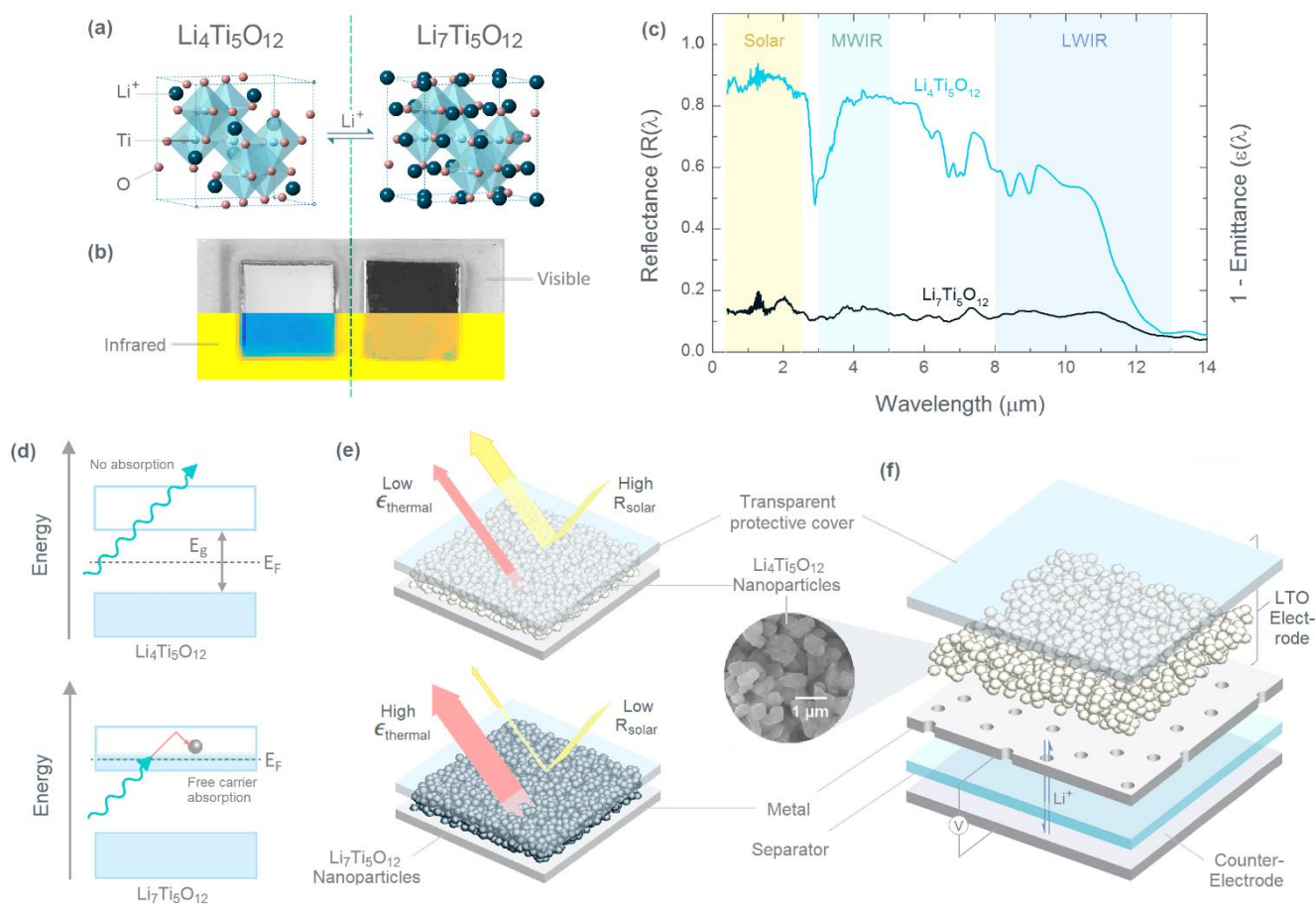

**Figure 1**. (a) Crystal structures of $Li_4Ti_5O_{12}$ and $Li_7Ti_5O_{12}$. (b) Photograph (top) and a LWIR thermograph (bottom) showing the different 'colors' of a $Li_{4+x}Ti_5O_{12}$ nanoparticle layer in the two states. The orange and blue colors indicate high and low LWIR emittances ($\epsilon_{LWIR}$) respectively. (c) Visible-to-infrared spectral reflectance $R(\lambda)$ of $Li_4Ti_5O_{12}$ and $Li_7Ti_5O_{12}$ nanoparticles (mass loading 2 mg cm$^{-2}$) deposited on Al foil and covered by barium fluoride ($BaF_2$). Solar, MWIR and LWIR ATWs are highlighted. (d) Schematic electron band-diagrams of the L and DL states, showing how a change in the electron energy distribution leads to (e) a super-broadband emittance and reflectance respectively. (f) Schematic of an electrochemical cell containing an 'outward-facing' LTO-based electrode, which consists of LTO nanoparticles deposited on a porous metal that allows the passage of $Li^+$.

The highly contrasting optical behavior of the L and DL states can be attributed to their respective electronic structures. In the DL state ($Li_4Ti_5O_{12}$), LTO is a wide band-gap semiconductor (Figure 1d), with a band-gap of ~3 eV.[11] Consequently, it has a low absorptance or emittance in the visible to LWIR wavelengths, and when nanostructured, can effectively back-scatter light and exhibit high reflectance (Figure 1e, upper panel). The L state ($Li_7Ti_5O_{12}$), on the other hand, is metallic (Figure 1d),[11a, 12] which causes $Li_7Ti_5O_{12}$ nanoparticles to act as a lossy, effective medium exhibiting high, broadband emittance in the infrared (Figure 1e, lower panel ).[13] This is consistent with FDTD simulation results based on Drude-Lorentz models of the two states (Supporting Information, Sections 2 and 3).[11] It is possible that polarons, which are often observed in $Li^+$-intercalated transition metal oxides,[6b, 14] contribute to the absorption in the visible wavelengths. Any contribution, however, is difficult to distinguish from the interactions between light and nanostructured plasmonic $Li_7Ti_5O_{12}$. The optical contrast between L and DL states disappears beyond $\lambda$ ~ 12 μm due to phonon modes that lead to high emittances for both L and DL states.[11c, 15] In addition, dips in $R(\lambda)$ at ~ 3, 7 and 8.5 μm for DL state arise from the IR absorption of poly(vinylidene fluoride) (PVdF) binder in the nanoparticle layer. It should be noted that the nanostructured LTO layer (Figure 1f) largely determines the optical contrast in the visible wavelengths, as unlike in bulk solids, scattering of light by the nanoparticles gives the DL state a matte white appearance and the L state a black color, regardless of the reflectance of the underlying metal. This is desirable for device design, since it allows metals with different visible reflectivities (e.g. aluminum, gold and copper) to be used as substrate, without significantly impacting the achievable optical tunability.

To quantify the broadband electrochromic performance of LTO on Al, the solar reflectance $R_{solar}$ and MWIR and LWIR thermal emittances $\epsilon_{MWIR}$ and $\epsilon_{LWIR}$ are calculated from $R(\lambda)$ (Supporting Information, Section 1) for L and DL states. High tunabilities of the reflectance or emittance – $\Delta R_{solar}$ ~ 0.74, $\Delta\epsilon_{MWIR}$ ~ 0.68 and $\Delta\epsilon_{LWIR}$ ~ 0.30 – are observed between states. $\Delta\epsilon_{LWIR}$ is comparable to or better than those of $WO_3$-based devices,[6a, 7] while $\Delta\epsilon_{MWIR}$ is on par with notable reports in the literature.[5, 8] Promisingly, $\Delta R_{solar}$ is higher than those exhibited by existing visible-to-infrared electrochromic devices known to the authors,[5, 7a] and together with the infrared $\Delta\epsilon$ values, makes LTO-based electrochromic devices (Figure 1f) highly attractive for thermal management and infrared camouflage.

**2.2 Optimization and Characterization of LTO Electrodes**

The super-broadband electrochromic behavior of LTO can be harnessed for thermal and optical applications by suitably designed electrochemical cells. In such a cell, the LTO-on-metal electrode faces 'outwards' for its electrochromism to be apparent (Figure 1f). A cover, which is transparent from the solar to LWIR wavelengths, is placed on top to protect the LTO from air. The metal contact that lies underneath the LTO is porous to facilitate the movement of Li$^+$ ions between the LTO and the counter-electrode. A proof-of-concept of this design is presented later in the paper.

The electrochromic performance of LTO depends on multiple factors, such as the type of cover, mass-loading of the LTO nanoparticle layer, radiative incidence or emittance angle ($\theta$), charging/discharging rate and number of electrochemical cycles. To provide a basis for device design, the authors investigate the effects of these parameters on the performance of LTO nanoparticles electrode-coated onto Al foil (**Figure 2**). First, the effect of the type of cover is investigated. As shown in in Figure 2a, a barium fluoride (BaF$_2$) protective cover offers greater transparency, while poly(ethene) (PE), due to its intrinsic emittance at $\lambda \sim 3.5$ and 7 μm (Supporting Information, Figure S3a), lowers $\Delta\epsilon_{MWIR}$ and $\Delta\epsilon_{LWIR}$ (0.59 and 0.24 compared to 0.68 and 0.30 for BaF$_2$). Due to its high transparency, BaF$_2$ is used for further investigations (Figure 2b, c and d). However, PE's lower cost and excellent flexibility make it more desirable for practical use.

Mass-loading of LTO on Al has a large impact on $R_{solar}$, $\epsilon_{MWIR}$ and $\epsilon_{LWIR}$, making it an important determinant of the performance of LTO electrodes. As evident from Figure 2b, DL LTO electrodes show an increase in $R_{solar}$ from 0.78 to 0.89 as mass loading increases from 0.3 to 4.4 mg cm$^{-2}$. This occurs because of an increased backscattering of light by the LTO nanoparticles. For the L state, $R_{solar}$ drops from 0.22 to 0.11 with mass loading due to an increase in the thickness of the absorptive LTO layer. Consequently, a maximum $\Delta R_{solar}$ of 0.78 is observed for the highest mass loading. For $\epsilon_{MWIR}$ and $\epsilon_{LWIR}$, an increase with mass loading is observed for both L and DL states. This is because the intrinsic emissivity of the LTO (small for DL state and large for L state) and the PVdF binder leads to higher $\epsilon$ as the amount of LTO increases. For the high mass loading of 4.4 mg cm$^{-2}$, even the DL state is quite emissive, causing $\Delta\epsilon_{MWIR}$ and $\Delta\epsilon_{LWIR}$ to drop. Peak values of $\Delta\epsilon_{MWIR} \sim 0.68$ and $\Delta\epsilon_{LWIR} \sim 0.30$ are observed for a mass loading of 2 mg cm$^{-2}$. $\Delta\epsilon$ is smaller in the LWIR due the phonon modes which raise $\epsilon_{LWIR}$ for both L and DL states. Since the 2 mg cm$^{-2}$ mass loading yields high values of $\Delta R_{solar}$ (0.74), $\Delta\epsilon_{MWIR}$ (0.68) and $\Delta\epsilon_{LWIR}$ (0.30) for super-broadband electrochromic applications, it is used henceforth, unless mentioned otherwise.

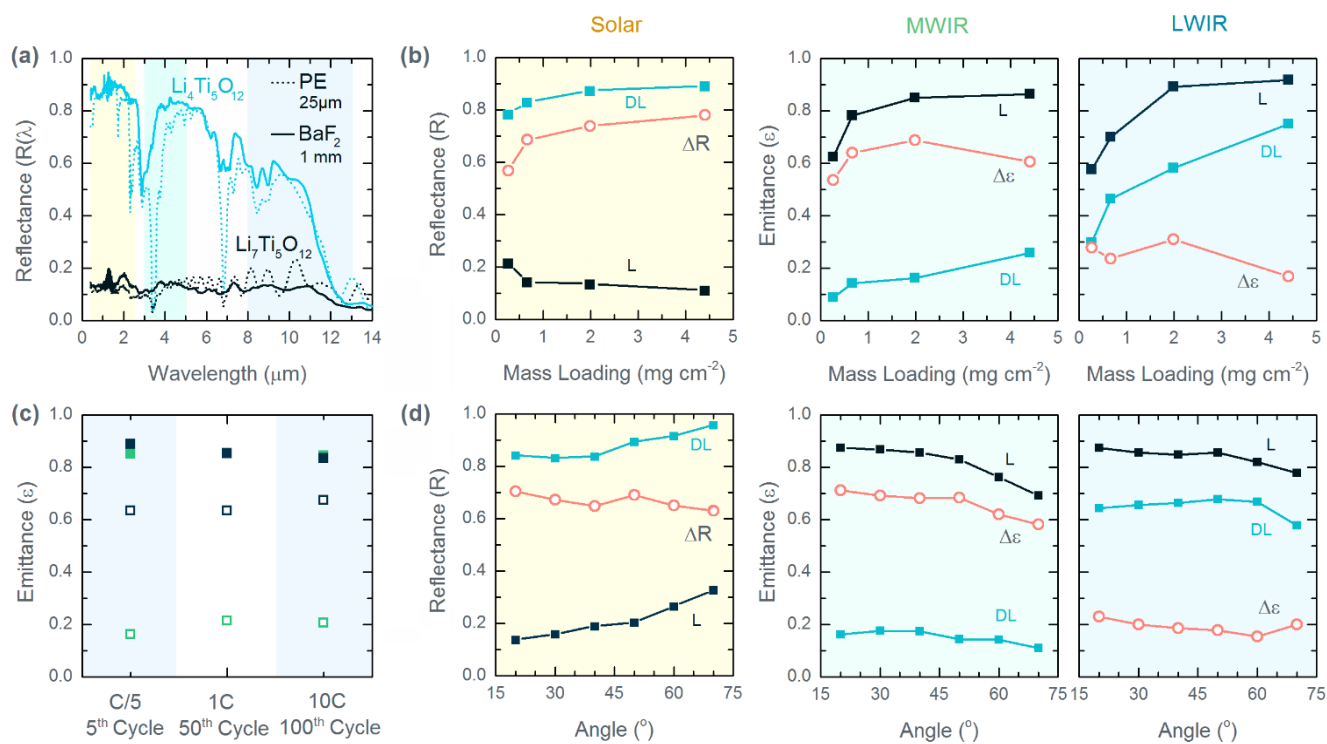

**Figure 2.** Variation of the emittance of the L and DL states of the LTO-based electrodes with (a) type of protective cover (polyethylene (PE) and Barium fluoride (BaF$_2$)), (b) mass loading, (c) electrochemical cycling and (d) angle. The background colors in b) and d) represent the solar, MWIR and LWIR atmospheric windows highlighted in (a). For (c), the green (■) and dark blue (■) colors indicate MWIR and LWIR emittances respectively, and for each color, the solid (■) and open (□) markers represent the L and DL states.

LTO also retains good optical performance during electrochemical cycling tests. For charge/discharge rates of C/5, 1C and 10C (1C= 175 mA g$^{-1}$), respective capacities of 168 mAh g$^{-1}$, 147 mAh g$^{-1}$ and 44 mAh g$^{-1}$ are observed (Supporting Information, Figure S4). Despite the drops in capacity, $\Delta\epsilon_{MWIR}$ and $\Delta\epsilon_{LWIR}$ remain quite steady compared to initial values (0.69 and 0.25 respectively), even after 50 cycles at 1C (0.64 and 0.22) 100 cycles at 10C (0.63 and 0.16) (Figure 2c). This indicates that a partial lithiation is enough to induce large $\Delta\epsilon_{MWIR}$ and $\Delta\epsilon_{LWIR}$, which would reduce switching time in electrochromic devices. The performance could potentially be enhanced by using smaller LTO nanoparticles or highly structured LTO, doping the nanoparticles to increase conductivity, compacting or sintering the LTO nanoparticle layer, and adding conductive indium tin oxide nanoparticles into it.[10, 16] Charge/discharge rates as high as 100 C has been achieved with LTO in earlier reports.[10] Significantly, a rate of 1C is suitable for switchable thermoregulation within solar timeframes (~1 h), while 10C approaches the rapid switching time (~1 min) required for camouflage. Therefore, the results show the promise of LTO-based electrochromic devices for such applications.

For practical use, an electrochromic device should also show optical tunability regardless of the viewing angle. For LTO on Al, angular measurements (Figure 2d) show that $\Delta R_{solar}$, $\Delta\epsilon_{MWIR}$ and $\Delta\epsilon_{LWIR}$ remain high from near-normal (20°) to near grazing (70°) angles ($\theta$) of emission or incidence. Except for Li$_4$Ti$_5$O$_{12}$ in the LWIR, $\epsilon(\theta)$ (=1- $R(\theta)$) decreases gradually with angle, which could be explained by both Fresnel reflection off the BaF$_2$ cover and Mie-scattering losses from the LTO nanoparticles at high angles. For Li$_4$Ti$_5$O$_{12}$ in the LWIR wavelengths, the optical behavior is compounded by the BaF$_2$ cover, which has high $\epsilon_{LWIR}$ at large angles (Supporting Information, Figure S3b). Consequently, the $\epsilon_{LWIR}$ of Li$_7$Ti$_5$O$_{12}$ stays near-constant and that for Li$_4$Ti$_5$O$_{12}$ rises gradually from 15° up to 50°. Overall, however, all the trends yield high $\Delta R$ or $\Delta\epsilon$ across the measured angles, implying a wide-angle behavior suited for thermal and optical applications.

## 2.3 Infrared-Camouflaging Capability of LTO

The high, wide-angle $\Delta\epsilon_{MWIR}$ and $\Delta\epsilon_{LWIR}$ of the LTO-based electrodes make them appealing for thermal camouflaging applications. To demonstrate the suitability of LTO for thermal camouflage, MWIR and LWIR thermographs of LTO on Al with BaF$_2$ and PE covers are presented in **Figure 3a**. As shown, in both the MWIR and LWIR bands, the non-emissive DL state appears colder, like the copper substrate, and the emissive L state appears hotter, like paper and glass. It is also observed that the difference in apparent temperature $T_{app,LTO}$ is higher in the MWIR than in the LWIR, and for the BaF$_2$-covered LTO on Al than for the PE covered one. These observations can be explained using the $\epsilon_{MWIR}$ and $\epsilon_{LWIR}$ values derived from Figures 1c and 2a and the following relation:

$$T_{app,LTO}^4 = \epsilon_{LTO}T_{LTO}^4 + R_{LTO}T_{env,rad}^4 \qquad (1),$$

where $T_{LTO}$ is the real temperature of the LTO-based electrode, $T_{env,rad}$ is the effective radiative temperature of the environment (Supporting Information, Section 6), $\epsilon_{LTO}$ is the emittance of the electrode, and $R_{LTO} = (1 - \epsilon_{LTO})$ is its reflectance (Figure 3b). As evident from Equation 1, in its emissive L state, LTO shows its true temperature, while in its reflective DL state, it mirrors the temperature of its cold environment. At higher real temperatures, $\epsilon_{LTO}$ carries a greater weight, making the $\Delta T_{app,LTO}$ between the states particularly prominent.

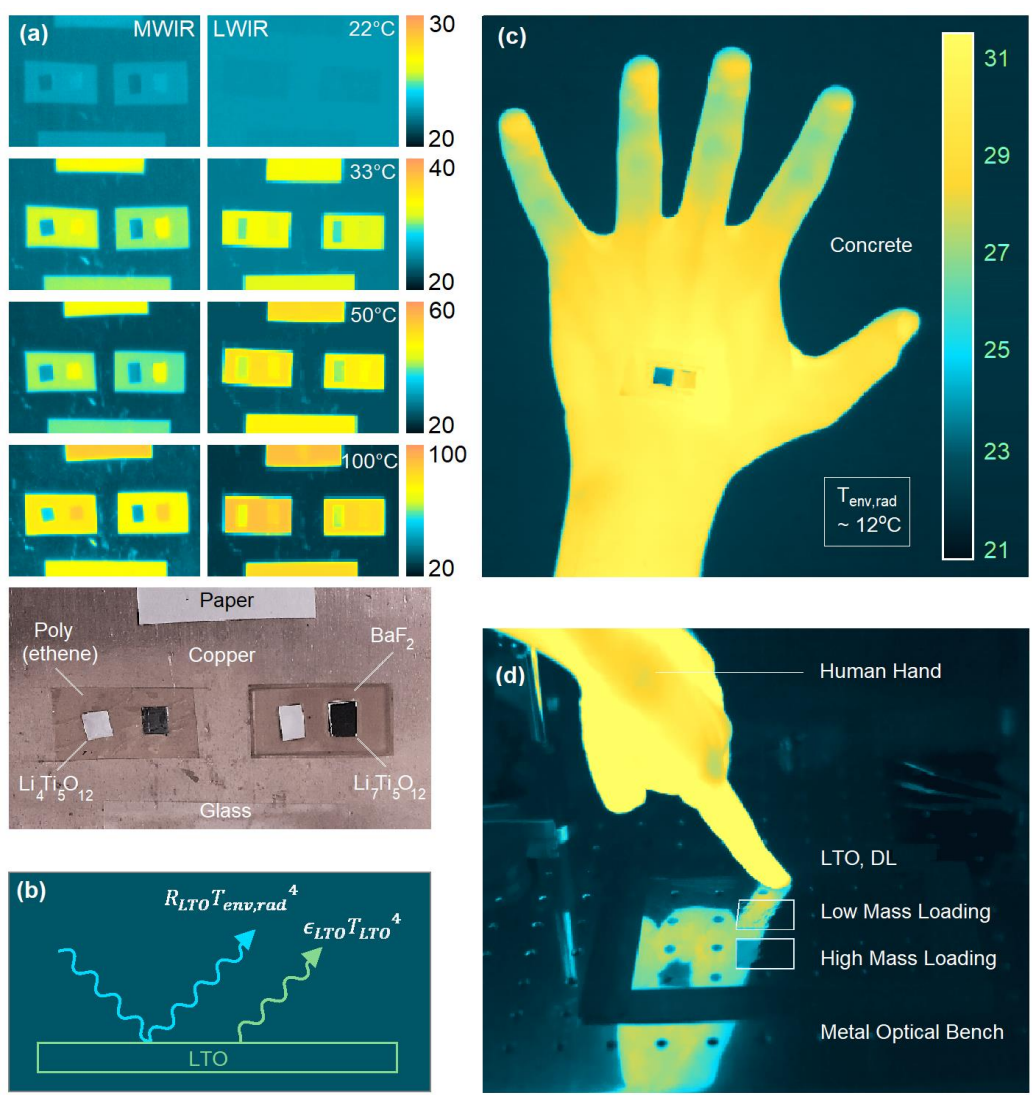

**Figure 3**. (a) MWIR and LWIR Thermographs of PE- and BaF$_2$-coated sets of LTO on Al, showing their behavior at room temperature (~22°C), on human skin (~31°C), and on hot machinery (50-100°C, e.g. vehicle surfaces). The photograph underneath shows the setup in the visible wavelengths. (b) Schematic showing the contribution of the environmental and intrinsic radiation to LTO's thermal signature. (c) LWIR thermograph of a hand, with PE-coated L and DL LTO, held against concrete. (d) LWIR thermograph showing specular reflections off PE-coated DL LTO with low (0.3 mg cm$^{-2}$) and high (4.4 mg cm$^{-2}$) mass loadings. Corresponding photographs for (c) and (d) are presented in the Supporting Information, Section 6.

The infrared electrochromic behavior implies that LTO can adjust $T_{app,LTO}$ to blend in with the surroundings. This is demonstrated in Figure 3c, where lithiated and delithiated LTO electrodes placed on a hand show LWIR temperatures (29.3°C) and (23.8°C) close to human skin (~31°C) and the concrete background (22.1°C) respectively. The camouflaging capability is expected to be even better in the MWIR, where $\Delta\epsilon$ is greater. The infrared tunability could be further optimized by tuning mass loading and extent of lithiation (Figures 2b, 3d and S3c). Furthermore, Figure 3d also shows that the LWIR reflections of LTO on Al is specular to certain extent. This means that LTO layers on Al mirror not only the environment's temperature, but also its thermal features. All these characteristics make LTO attractive for thermal camouflage.

**2.4 Solar Heating and Radiative Cooling Capabilities of LTO**

The super-broadband electrochromism of LTO also makes it promising for radiative thermoregulation. In this regard, the solar and LWIR wavelengths are important. For instance, a high $R_{solar}$ can lead to a lowered heating by the sun, while a high $\epsilon_{LWIR}$ can lead to a higher radiative heat loss from objects in everyday situations.[17] For LTO nanoparticles on Al, $\Delta R_{solar}$ is greater than accompanying $\Delta\epsilon_{LWIR}$. Furthermore, solar intensities (~ 900 W m$^{-2}$ at noon) usually exceed LWIR irradiances (< 300 W m$^{-2}$) from everyday objects (temperature < 75°C) (Supporting Information, Section 7). Therefore, during the day, heating or cooling can be achieved by tuning $R_{solar}$. For example, under a solar intensity $I$ ~ 788 Wm$^{-2}$, black L and white DL states of LTO (Figure 4a) reach significantly different temperatures of 23°C and 5°C relative to the ambient air (Figure 4b-c) respectively. At night, $\epsilon_{LWIR}$ can be tuned to control radiative heat loss. As demonstrated, the L state's a higher $\epsilon_{LWIR}$ enables it to reach a relative temperature of -2.7°C, while the DL state attains a relative temperature of -2°C (Figure 4c). The optical performance can be further modified by placing additional layers over the LTO. For instance, when a few layers of solar-reflective, LWIR-transmissive nanoporous polyethylene separator is used, both L and DL states reach sub-ambient temperatures (-0.7 and -3.9°C relative to the surrounding air respectively) under $I \sim 805$ Wm$^{-2}$ (Figure 4c and Supporting Information, Section 7). LTO-based electrochromic devices thus have both radiative heating and cooling capabilities.

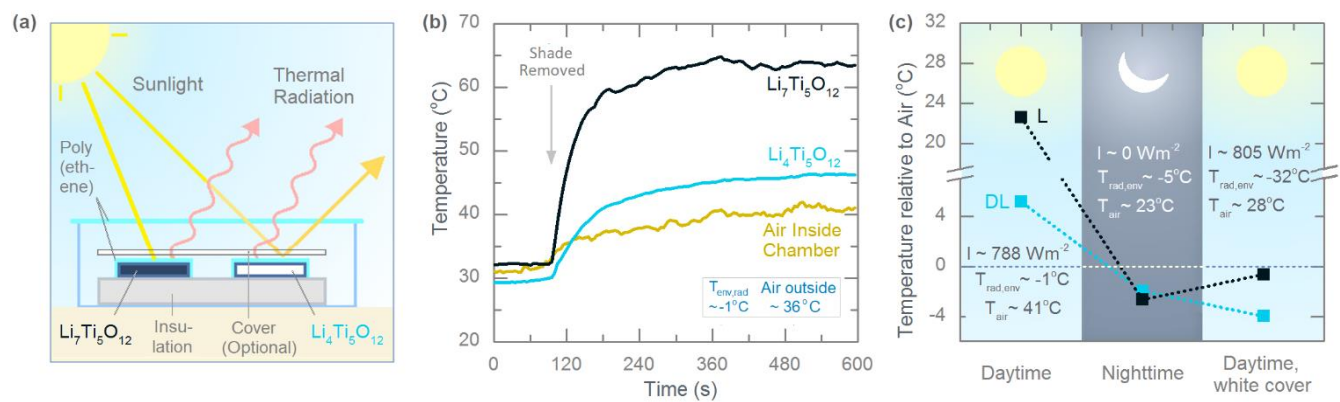

**Figure 4.** (a) Schematic of the experimental setup for the thermoregulation tests, showing the PE-coated L and DL LTO on Al in a chamber with a PE top-cover. Photographs are presented in the Supporting Information, Section 7. (b) Temperature-time plots of the samples and the ambient air under sunlight. (c) Equilibrium temperatures relative to the ambient air of the L (Li$_7$Ti$_5$O$_{12}$) and DL (Li$_4$Ti$_5$O$_{12}$) LTO on Al under sunlight, at night, and under sunlight with a white nanoporous PE cover.

The tuneable $R_{solar}$ and $\epsilon_{LWIR}$ of LTO make it attractive for thermal management both in space and on earth. For space applications, Chandrasekhar et. al. have described a conducting-polymer-based electrochromic skin with an excellent, super-broadband $\Delta\epsilon$ ($\sim$ 0.5).[5] However its $\Delta R_{solar}$ is low ($<$ 0.15), and since sunlight experienced by satellites and spacecraft around the earth is particularly strong ($\sim$1350 Wm$^{-2}$), that could limit thermoregulation capability. In this regard, LTO's large $\Delta R_{solar}$ ($>$ 0.7) could enable a greater control of solar absorptance when it is sunlit, while its MWIR-to-LWIR $\Delta\epsilon$ would allow a tunable radiative heat loss when it is shaded. On earth, potential applications include the thermoregulation of smart-roofs, storage tanks and warehouses which, depending on the time of the day or the year, could be radiatively heated or cooled to desired temperatures.[18] As shown in the Supporting information, Section 7, relative to a concrete surface with $R_{solar} \sim 0.5$,[19] LTO-based electrodes could achieve increased radiative heat losses or gains of $>$ 1.5 kWh m$^{-2}$/day. The value corresponds to energy savings that are both economically and environmentally significant,[20] and underlines LTO's potential for energy-efficient, eco-friendly thermoregulation.

## 2.5 Proof of Concept Electrochemical Cell

Besides the super-broadband optical tunability of LTO, device design is also important for a functional electrochromic device. Effective designs for infrared electrochromism with metal substrates have previously been reported in the literature.[5-8] Here, a proof-of-concept of the design shown in Figure 1f is demonstrated in Figure S8 of the Supporting Information. In the design, LTO nanoparticles, deposited on a steel mesh and facing outwards from the electrochromic cell, is the cathode. LiPF$_6$ in ethyl/diethyl carbonate is used as the organic liquid electrolyte and PE is used as the cover material. Successful switching between L and DL states is observed. The effect of organic liquid electrolytes on the performance of the LTO electrode is also investigated, and is found to be small – as shown in Figure S9 and Table S4 of the Supporting Information, $\Delta\epsilon_{MWIR}$ decreases from 0.68 to 0.62 and $\Delta\epsilon_{MWIR}$ decreases from 0.24 to 0.23 when electrolyte is present. The results show that liquid-electrolyte-based designs are viable and promising for future exploration. Another possible design would be a solid-state device containing a LTO electrode integrated with a solid electrolyte. Such a device would not have leakage issues, and be operable at temperatures beyond the range of existing aqueous designs.[5, 7a, 8, 21]. While this study focuses on material properties, further work on in-situ electrochromic switching of Li$_4$Ti$_5$O$_{12}$ remains to be done for device applications.

## 3. Conclusion

In summary, the authors demonstrate Li$_4$Ti$_5$O$_{12}$ (LTO), which changes from a wide band-gap semiconductor to metal upon lithium intercalation, as a potential visible-to-infrared super-broadband electrochromic material for optical and thermal management. Lithiated (Li$_7$Ti$_5$O$_{12}$) and delithiated (Li$_4$Ti$_5$O$_{12}$) states of LTO- nanoparticles on Al exhibit high $\Delta\epsilon$ or $\Delta R$ of 0.74, 0.68 and 0.30 in the solar, MWIR and LWIR wavelengths respectively. The optical performance holds from near-normal to grazing incidence angles of light, and is tunable by changing the mass-loading and degree of lithiation of LTO. Moreover, LTO show steady electrochromic performances under different electrochemical cycling tests. The range of $\epsilon$ attained by LTO on Al allows them to camouflage themselves by reflecting their environment's temperature and thermal features, and makes them appealing for thermal camouflaging applications. Furthermore, under different daytime and nighttime skies, LTO shows both solar heating and radiative cooling capabilities. The large tunable temperature range (18°C) and sub-ambient (by ~4°C) temperatures attained by LTO under sunlight make it attractive for energy-efficient thermoregulation in space and terrestrial settings. $Li_4Ti_5O_{12}$ is thus promising for infrared camouflage and thermoregulation.

## 4. Experimental Section

*Fabrication of the LTO-based electrodes for device optimization:* $Li_4Ti_5O_{12}$ nanoparticles from Hydroquebec were ground with poly(vinylidene fluoride) powder (Arkema Kynar HSV900) in a 98:2 weight ratio, and then added to 1-methyl-2-pyrolidinone (NMP) to form slurries. Applicators were then used to make $Li_4Ti_5O_{12}$ nanoparticle layers on aluminum foil. Depending on their use, the layers were coated with $BaF_2$ or PE straightaway or after electrochemical processing.

*Cell assembly, electrochemical processing and disassembly:* LTO-based electrodes were put in pouch cells in a Li | electrolyte (Gotion LP40) | $Li_4Ti_5O_{12}$ configuration in an inert atmosphere. The cells were then charged under constant-current-constant-voltage regimes to yield $Li_7Ti_5O_{12}$, or cycled at different rates and for different number of cycles, using a potentiostat (Bio-logic, Model VMP3). Afterwards, the cells were disassembled, and the LTO-based electrodes structures were cleaned in diethylene carbonate, heated dry, and then covered with $BaF_2$ or PE.

*Optical characterization:* Spectral reflectance $R(T,\lambda)$ of LTO nanoparticles on Al was separately measured in the visible to near-infrared (0.41-1.05 μm) and near-infrared to mid-infrared (1.06-14 μm) wavelength ranges. For the first case, a supercontinuum laser (SuperK Extreme, NKT Photonics) coupled to a tunable filter (Fianium LLTF contrast) was used to shine specific wavelengths into an integrating sphere (Model IS200, Thorlabs). The samples were then inserted into the integrating sphere to intercept the light at angles between 15° and 70°, and had their reflectances measured at 5 nm wavelength intervals by a silicon detector. A patch of the standard white material of the integrating sphere was used as reference. For the 1.06-14 μm range, a Fourier Transform Infrared spectrometer (Vertex 70v, Bruker), gold-coated aluminum foil references, and a gold integrating sphere (Model 4P-GPS-020-SL, Labsphere) along with a mercury-cadmium telluride detector were likewise used. The obtained spectra were then patched, and $\bar{\epsilon}(\lambda_1, \lambda_2)$ and $\bar{R}(\lambda_1, \lambda_2)$ were calculated from the resultant spectrum using Equation 1 of the Supporting Information.

*Thermography to demonstrate infrared camouflaging:* MWIR and LWIR thermographs of samples were taken using FLIR SC6000 and T640 thermal cameras respectively.

*Solar-thermoregulation test:* Lithiated and delithiated samples of LTO on Al were placed on a thermally insulating foam, and then in a white, open-top box with a polyethene cover to reduce convection and allow transmission of solar and thermal radiation. The setup was then placed on a horizontal rooftop or an open courtyard at 40.8093°N, -73.9597°E and allowed to reach steady-state under sunlight. Incident solar intensity was calculated using a thermopile sensor (type 3A, item 7Z02621, Ophir Photonics) and angular position of the Sun. The FLIR T640 LWIR thermal camera was used to measure the radiative temperature of the environment. Temperatures of the samples, the air inside the box and the air outside was measured with thermocouples. More information is provided in the Supporting Information, Section 7.

*SEM characterization:* $Li_4Ti_5O_{12}$ nanoparticles on Al foil were imaged with a Zeiss Sigma VP scanning electron microscope.

**Supporting Information**

Supporting Information is available from the Wiley Online Library or from the authors.

**Acknowledgements**

The work was supported by startup funding from Columbia University, the NSF MRSEC program through Columbia in the Center for Precision Assembly of Superstratic and Superatomic Solids (DMR-1420634), AFOSR MURI (Multidisciplinary University Research Initiative) program (grant # FA9550-14-1-0389), and AFOSR DURIP (Defense University Research Instrumentation Program) (grant # FA9550-16-1-0322). Optimization and synthesis of LTO nanoparticles was supported by Hydro Québec. The authors would also like to thank Adam Overvig of the Department of Applied Physics at Columbia University for his help with this study, and Lin Zhao and Prof. Evelyn Wang of the Department of Mechanical Engineering at Massachusetts Institute of Technology for use of their MWIR camera.

Received: ((will be filled in by the editorial staff))
Revised: ((will be filled in by the editorial staff))
Published online: ((will be filled in by the editorial staff))

**References**

[1]  a) D. R. Rosseinsky, R. J. Mortimer, *Advanced Materials* **2001**, 13, 783; b) D. M. DeLongchamp, P. T. Hammond, *Chemistry of Materials* **2004**, 16, 4799; c) A. Llordes, G. Garcia, J. Gazquez, D. J. Milliron, *Nature* **2013**, 500, 323; d) E. L. Runnerstrom, A. Llordes, S. D. Lounis, D. J. Milliron, *Chemical Communications* **2014**, 50, 10555; e) Preionic Technologies, Electrochromic Displays, http://www.prelonic.com/products/electrochromic-displays.html, accessed: Octorber, 2017; f) Gentex Corporation, Aircraft Windows, http://www.gentex.com/aerospace/aircraft-windows, accessed: October, 2017.

[2]  H. Yu, S. Shao, L. Yan, H. Meng, Y. He, C. Yao, P. Xu, X. Zhang, W. Hu, W. Huang, *Journal of Materials Chemistry C* **2016**, 4, 2269.

[3]  a) D. Rosseinsky, P. Monk, R. Mortimer, *Electrochromism and Electrochromic Devices*, Cambridge University Press, Cambridge **2007**; b) S. K. Deb, *Appl. Opt.* **1969**, 8, 192; c) T. Kobayashi, H. Yoneyama, H. Tamura, *Journal of Electroanalytical Chemistry and Interfacial Electrochemistry* **1984**, 161, 419; d) C. J. Barile, D. J. Slotcavage, M. D. McGehee, *Chemistry of Materials* **2016**, 28, 1439; e) P. R. Somani, S. Radhakrishnan, *Materials Chemistry and Physics* **2003**, 77, 117; f) R. J. Mortimer, *Annual Review of Materials Research* **2011**, 41, 241; g) P. Shi, C. M. Amb, E. P. Knott, E. J. Thompson, D. Y. Liu, J. Mei, A. L. Dyer, J. R. Reynolds, *Advanced Materials* **2010**, 22, 4949; h) S. H. Lee, R. Deshpande, P. A. Parilla, K. M. Jones, B. To, A. H. Mahan, A. C. Dillon, *Advanced Materials* **2006**, 18, 763; i) A. A. Argun, P.-H. Aubert, B.

C. Thompson, I. Schwendeman, C. L. Gaupp, J. Hwang, N. J. Pinto, D. B. Tanner, A. G. MacDiarmid, J. R. Reynolds, *Chemistry of Materials* **2004**, 16, 4401; j) G. Sonmez, H. B. Sonmez, C. K. F. Shen, F. Wudl, *Advanced Materials* **2004**, 16, 1905; k) L. Sicard, D. Navarathne, T. Skalski, W. G. Skene, *Advanced Functional Materials* **2013**, 23, 3549; l) R. T. Wen, C. G. Granqvist, G. A. Niklasson, *Advanced Functional Materials* **2015**, 25, 3359; m) W. Zhen, Z. Qingzhu, C. Shan, C. Zhigang, Z. Jinxiong, Y. Mei, Z. Zuhui, Z. Sha, Y. Xuwen, G. Fengxia, Z. Zhigang, *Advanced Optical Materials* **2017**, 5, 1700194; n) C. Shan, G. Fengxia, Z. Zhigang, *Advanced Materials* **2016**, 28, 10518.

[4] A. Rogalski, K. Chrzanowski, *Opto-Electronics Review* **2002**, 10, 111.
[5] P. Chandrasekhar, B. J. Zay, D. Lawrence, E. Caldwell, R. Sheth, R. Stephan, J. Cornwell, *Journal of Applied Polymer Science* **2014**, 131.
[6] a) H. Li, K. Xie, Y. Pan, M. Yao, C. Xin, *Synthetic Metals* **2009**, 159, 1386; b) A. Hjelm, C. G. Granqvist, J. M. Wills, *Physical Review B* **1996**, 54, 2436.
[7] a) A. Bessière, C. Marcel, M. Morcrette, J. M. Tarascon, V. Lucas, B. Viana, N. Baffier, *Journal of Applied Physics* **2002**, 91, 1589; b) E. B. Franke, C. L. Trimble, J. S. Hale, M. Schubert, J. A. Woollam, *Journal of Applied Physics* **2000**, 88, 5777.
[8] A. Rougier, K. Sauvet, L. Sauques, *Ionics* **2008**, 14, 99.
[9] M. S. Song, A. Benayad, Y.-M. Choi, K.-S. Park, *Chemical Communications* **2012**, 48, 516.
[10] X. Sun, P. V. Radovanovic, B. Cui, *New Journal of Chemistry* **2015**, 39, 38; b) T. Linkai, H. Yan‐Bing, W. Chao, W. Shuan, W. Marnix, L. Baohua, Y. Quan‐Hong, K. Feiyu, *Advanced Science* **2017**, 4, 1600311; c) Y.-S. Lin, M.-C. Tsai, J.-G. Duh, *Journal of Power Sources* **2012**, 214, 314; d) L. Shen, C. Yuan, H. Luo, X. Zhang, S. Yang, X. Lu, *Nanoscale* **2011**, 3, 572; e) A. Mauger, C. Julien, *Nanomaterials* **2015**, 5, 2279.
[11] a) C. Y. Ouyang, Z. Y. Zhong, M. S. Lei, *Electrochemistry Communications* **2007**, 9, 1107; b) Ö. Soner, Ş. Volkan, P. Suat, K. Şadan, *Journal of Physics D: Applied Physics* **2016**, 49, 105303; c) M. W. Raja, S. Mahanty, M. Kundu, R. N. Basu, *Journal of Alloys and Compounds* **2009**, 468, 258.
[12] M. G. Verde, L. Baggetto, N. Balke, G. M. Veith, J. K. Seo, Z. Wang, Y. S. Meng, *ACS Nano* **2016**, 10, 4312.
[13] a) D. Young, A. Ransil, R. Amin, Z. Li, Y.-M. Chiang, *Advanced Energy Materials* **2013**, 3, 1125; b) J. Mandal, D. Wang, A. C. Overvig, N. N. Shi, D. Paley, A. Zangiabadi, Q. Cheng, K. Barmak, N. Yu, Y. Yang, *Advanced Materials* **2017**, 29, 1702156.
[14] a) C. G. Granqvist, *Applied Physics A* **1993**, 57, 3; b) G. A. Niklasson, C. G. Granqvist, *Journal of Materials Chemistry* **2007**, 17, 127.
[15] P. P. Prosini, R. Mancini, L. Petrucci, V. Contini, P. Villano, *Solid State Ionics* **2001**, 144, 185.
[16] a) Y. Yang, S. Jeong, L. Hu, H. Wu, S. W. Lee, Y. Cui, *Proceedings of the National Academy of Sciences* **2011**, 108, 13013. b) C. Lim, B. Yan, H. Kang, Z. Song, W. C. Lee, V. De Andrade, F. De Carlo, L. Yin, Y. Kim, L. Zhu, *Journal of Power Sources* **2016**, 328, 46.
[17] a) A. P. Raman, M. A. Anoma, L. Zhu, E. Rephaeli, S. Fan, *Nature* **2014**, 515, 540; b) A. Srinivasan, B. Czapla, J. Mayo, A. Narayanaswamy, *Applied Physics Letters* **2016**, 109, 061905.
[18] a) S. Soper, in *The Morning Call*, The Morning Call, Allentown, PA 2011; b) E Source Companies LLC, Improving Energy Efficiency in Warehouses, http://bea.touchstoneenergy.com/sites/beabea/files/PDF/Sector/Warehouses.pdf, accessed: October, 2017; c) Pentair Thermal, Tank Heating: Store Industrial Liquids at the Right Temperature and Minimize Energy Consumption, https://www.pentairthermal.com/application/tank-heating/industrial-facilities/index.aspx?seg=industrial, accessed: October, 2017.
[19] M. Marceau, M. VanGeem, *Concrete International* **2008**, 30, 08.
[20] a) U. S. G. S. Administration, *The Benefits and Challenges of Green Roofs on Public and Commercial Buildings.pdf*, United States General Services Administration, Washington, DC **2011**; b) U. E. I. Administration, State Electricity Profiles, https://www.eia.gov/electricity/state/, accessed.
[21] Y. Kato, S. Hori, T. Saito, K. Suzuki, M. Hirayama, A. Mitsui, M. Yonemura, H. Iba, R. Kanno, **2016**, 1, 16030.